\documentstyle[12pt,epsfig]{article}
\def\bge{\begin{equation}}
\def\ene{\end{equation}}
\def\bg{\begin{eqnarray}}
\def\en{\end{eqnarray}}
\def\nn{\nonumber}

\def\S0{{\Sigma^0}}

\def\kbar{\bar{K}}
\def\k0bar{\bar{K}^0}
\def\qqbar{q \bar{q}}

\def\ubar{\bar{u}}
\def\dbar{\bar{d}}
\def\vecr{\vec{r}}
\def\veck{\vec{k}}
\def\e{\epsilon}
\begin{document}
\begin{titlepage}
\title{In-medium kaon and antikaon properties \\
in the quark-meson coupling model}
\author{
K. Tsushima$^1$~\thanks{ktsushim@physics.adelaide.edu.au}~,
K. Saito$^2$~\thanks{ksaito@nucl.phys.tohoku.ac.jp}, 
A. W. Thomas$^1$~\thanks{athomas@physics.adelaide.edu.au}~and 
S. V. Wright$^1$~\thanks{swright@physics.adelaide.edu.au} \\
{\small $^1$Department of Physics and Mathematical Physics} \\
{\small and Special Research Center for the Subatomic Structure of Matter,} \\
{\small University of Adelaide, SA 5005, Australia} \\
{ $^2$\small Physics Division, Tohoku College of Pharmacy} \\
{\small Sendai 981, Japan} }
\maketitle
\vspace{-10cm}
\hfill ADP-97-50/T277
\vspace{10cm}
\begin{abstract}
The properties of the kaon, $K$, and antikaon, $\kbar$, in nuclear medium  
are studied in the quark-meson coupling (QMC) model. Employing  
a constituent quark-antiquark (MIT bag model) picture,  
their excitation energies in a nuclear medium at zero momentum
are calculated within mean field approximation.
The scalar, and the vector mesons  
are assumed to couple directly to the nonstrange quarks 
and antiquarks in the $K$ and $\kbar$ mesons. It is demonstrated that  
the $\rho$ meson induces different mean field potentials 
for each member of the isodoublets, $K$ and $\kbar$, 
when they are embedded in asymmetric nuclear matter.
Furthermore, it is also shown that this $\rho$ meson potential 
is repulsive for the $K^-$ meson in matter with a neutron excess, and  
renders $K^-$ condensation less likely to occur. 
\end{abstract}
\end{titlepage}
%

The study of the properties of the kaon, $K$, and antikaon, $\kbar$, 
in a dense nuclear medium is   
one of the most exciting new directions in nuclear 
physics. Stimulated by the pioneering work of Kaplan and 
Nelson~\cite{kap}, intensive work has been performed about 
the possibility of the $K^-$ meson condensation in 
a dense nuclear medium, and its effect on the properties of 
neutron stars~\cite{kk1} -~\cite{dai}.
In addition, many investigations have been 
made~\cite{brown2,li} concerning 
$K N$ and $\kbar N$ interactions~\cite{siegel} -~\cite{lut2}, and 
strangeness production in heavy ion 
collisions~\cite{li,kkh,ko,liflow,koch,bra}, with a particular emphasis  
on the medium modification of the kaon and antikaon properties. 

Although $K$ and $\kbar$ mesons are Goldstone bosons in the chiral limit, 
they are also expected to reveal a  
quark-antiquark ($\qqbar$) structure to some extent, 
because their mass is relatively heavy compared to that of the pion.
Indeed, the naive constituent quark model has proven  
quite successful for studying the properties of $K$ and $\kbar$ mesons 
in free space~\cite{deg}.

However, not so many investigations have been performed on 
the properties of $K$ and $\kbar$ mesons in nuclear medium with  
explicit quark degrees of freedom~\cite{lut1}.  
One of the main reasons for this is that there has been no appropriate 
model until recently, which can simultaneously describe the 
properties of the nuclear medium  
(finite nuclei), as well as hadron properties, based on quark degrees  
of freedom.  It now seems possible to do this because of the recent  
development of the quark-meson coupling (QMC) model, which 
was initiated by Guichon~\cite{gui}.
This model has been successfully applied to investigate the properties of
infinite nuclear matter~\cite{gui} -~\cite{pan} and
finite nuclei~\cite{finite1} -~\cite{hyper}, with some extension
to incorporate the self-consistent variation of 
the meson masses~\cite{finite2}. 
Although the QMC model may be regarded as a general 
extension of Quantum Hadrodynamics 
(QHD)~\cite{qhd}, the difference between the two models probably 
becomes distinctively clear when one investigates the properties of 
mesons in nuclear medium, where their quark structure  
plays a vital role.


Let the mean values of the meson fields, the scalar, 
the time component of the vector isoscalar, and the time  
component of the vector isovector in the third direction in isospin,  
be $\sigma$, $\omega$ and $b$, respectively,  
in a uniformly distributed nuclear medium. 
Here, we assume that the $K$ and $\kbar$ mesons are described by 
the static spherical MIT bag,  
as are the nucleons. We also suppose that the $\sigma$, 
$\omega$ and $\rho$ mesons only interact directly with the nonstrange  
quarks and antiquarks in the $K$ and $\kbar$ mesons. 
The Dirac equations for the corresponding  
wave functions for up and down quarks are given by:
\bge
\left[ i \gamma \cdot \partial - (m_q - V_\sigma) \mp \gamma^0 
\left( V_\omega + \frac{1}{2} V_\rho \right) \right]
\left( \begin{array}{c} u\\ \ubar\\ \end{array} \right) = 0,
\label{diracu}
\ene
\bge
\left[ i \gamma \cdot \partial - (m_q - V_\sigma) \mp \gamma^0 
\left( V_\omega - \frac{1}{2} V_\rho \right) \right]
\left( \begin{array}{c} d\\ \dbar\\ \end{array} \right) = 0, 
\label{diracd}
\ene
\label{diracequation}
where $V_\sigma = g^q_\sigma \sigma, V_\omega = g^q_\omega \omega$ and 
$V_\rho = g^q_\rho b$ 
with $g^q_\sigma, g^q_\omega$ and $g^q_\rho$ 
being, respectively, the corresponding quark and meson coupling constants. 
Here we assume SU(2) symmetry. Thus, the current masses 
for the quarks and antiquarks follow the relation, 
$m_q \equiv m_u = m_d = m_{\ubar} = m_{\dbar}$. 
The normalized, static solution 
for the ground state for a nonstrange 
quark or antiquark in the kaon or antikaon 
may be written as:
\bge
\psi_i (\vecr) = N_i e^{- i \epsilon_i t / R_K^\star} \phi_i (\vecr), 
\qquad {\rm for}\;\; i = u, \ubar, d, \dbar,
\label{wavefunction}
\ene
where $N_i$ and $\phi_i (\vecr)$ are respectively the normalization factor
and the corresponding spatial part of the wave function~\cite{tsushima}. 
The bag radius in medium, $R_K^\star$, which depends on the hadron species 
in which the quarks belong, will be determined self-consistently  
through Eqs.~(\ref{mass}) and~(\ref{equil})    
similarly to those for the octet baryons~\cite{finite1,finite2,hyper}. 
The quark eigenenergies in units of $1/R_K^\star$, $\epsilon_i\;
(i = u, \ubar, d, \dbar)$ in Eq. (\ref{wavefunction}), are given by 
\bge
\left( \begin{array}{c} \e_u \\ \e_{\ubar} \end{array} \right)
= \Omega^\star \pm R_K^\star \left( 
V_\omega + \frac{1}{2} V_\rho \right) 
\quad {\rm and}\quad
\left( \begin{array}{c} \e_d \\ \e_{\dbar} \end{array} \right)
= \Omega^\star \pm R_K^\star \left( 
V_\omega - \frac{1}{2} V_\rho \right), 
\label{quarkpot}
\ene
where $\Omega^\star = \sqrt{ x^{\star 2} + (R_K^\star m^\star_q)^2}$, with 
$m^\star_q = m_q - g^q_\sigma \sigma$. The quark eigenfrequency 
in medium, $x^\star$,  
is determined by the usual, linear boundary condition~\cite{finite0,tsushima}.
Then the excitation energies for the $K$ and $\kbar$ mesons with zero  
momenta, $\omega_i\; (i = K^+, K^-, K^0, \k0bar)$, are given by
\bge
\left( \begin{array}{c} \omega_{K^+} \\ \omega_{K^-} \end{array} \right)
= m^\star_K \pm \left( V_\omega + \frac{1}{2} V_\rho \right)
\quad {\rm and}\quad
\left( \begin{array}{c} \omega_{K^0} \\ \omega_{\k0bar} \end{array} \right)
= m^\star_K \pm \left( V_\omega - \frac{1}{2} V_\rho \right), 
\label{kaonpot}
\ene
where, the effective mass of the $K$ and $\kbar$ mesons, 
$m^\star_K = m^\star_{\kbar}$, is calculated using the MIT bag model:
\bg
& &m^\star_K = \frac{\Omega^\star + \Omega_s - z_K}{R^\star_K}
+ {4\over 3}\pi R^{\star 3}_K B, \label{mass}\\
& &\left.{\partial m_K^\star
\over \partial R}\right|_{R = R^\star_K} = 0, \label{equil}
\en
with the strange-quark energy in units of $1/R^\star_K$, 
$\Omega_s = \sqrt{x_s^2 + (R^\star_K m_s)^2}$, and $z_K$ in 
Eq.~(\ref{mass}) parametrizes the sum of the center-of-mass and gluon 
fluctuation effects. 

After self-consistent calculation, the effective mass of the $K$ and 
$\kbar$ mesons, $m^\star_K$, can be parametrized
in the applied scalar field~\cite{finite0,finite1,finite2,hyper}: 
\bge
m^\star_K \equiv m_K - g^K_\sigma (\sigma) \sigma 
\simeq m_K - \frac{1}{3} g^N_\sigma 
\left[ 1 - \frac{a_K}{2} (g^N_\sigma \sigma) \right] \sigma,
\label{appmass}
\ene
where $g^N_\sigma$ 
is the nucleon and $\sigma$ meson coupling constant in 
free space ($\sigma = 0$)~\cite{finite0,hyper}, and the relation, 
$g^K_\sigma = \frac{1}{3} g^N_\sigma$, involves an error 
less than 0.5 \%~\cite{finite1,finite2,hyper}.
(The quantity, $a_K$, in Eq.~(\ref{appmass}) is found to be
$a_K = 6.6 \times 10^{-4}$ MeV$^{-1}$.)
In this study, we chose the values, $m_u = m_d = 5$ MeV and $m_s = 250$ MeV, 
for the current quark masses, and $R_N = 0.8$ fm 
for the bag radius of the nucleon in free space. 
Other inputs and parameters used,
and some of the quantities calculated in the present study, are listed
in Table~\ref{parameters}.
(Note that $r_q$ is the root-mean-square
(r.m.s.) charge radius calculated using the MIT bag model
wave functions obtained by solving the set of equations including
the strange quark.)
We stress that while the model has a number of parameters, only three of them, 
$g^q_\sigma$, $g^q_\omega$ 
and $g^q_\rho$, are adjusted to fit nuclear data -- namely the saturation 
energy and density of symmetric nuclear matter and the bulk symmetry energy.
None of the results for nuclear properties depend strongly on the choice
of the other parameters -- for example, the
relatively weak dependence of the final results on the chosen values
of the current quark mass and bag radius is shown explicitly in
Refs.~\cite{finite0,finite1}.
%
\begin{table}[htbp]
\begin{center}
\caption{
Inputs, parameters and some of the quantities calculated in the present 
study. The quantities with star, $^\star$, are those quantities calculated  
at normal nuclear density, $\rho_0 = 0.15$ fm$^{-3}$. 
The values for the bag constant, and current 
quark masses are respectively, $B = (170.0 {\rm MeV})^4$, and 
$m_u = m_d = 5$ MeV and $m_s = 250$ MeV. 
}
\label{parameters}
\vspace{1em}
\begin{tabular}[hbtp]{|l|lllllll|}
\hline
&$m$ (MeV) &$z_{N,K}$ &$R$ (fm) &$r_q$ (fm) 
&$m^\star$ (MeV)&$R^\star$ (fm)&$r^\star_q$ (fm)\\
\hline
N &939.0 (input) &3.295 &0.800 (input) &0.582 &754.6 &0.786 &0.594 \\
K &493.7 (input) &3.295 &0.574         &0.412 &430.5 &0.572 &0.418 \\
\hline
\end{tabular}
\end{center}
\end{table}
%
%

Boosting the $K$ and $\kbar$ bags in a uniformly distributed vector field, 
we can find the dispersion relation for the $K$ and $\kbar$ mesons moving  
with momentum $\vec{k}$ as in Ref.~\cite{kkh}:
\bge
\left( \begin{array}{c} \omega_{K^+}(\veck) \\ 
\omega_{K^-}(\veck) \end{array} \right) = 
\sqrt{m^\star_K + \vec{k}^2} \pm \left( V_\omega + \frac{1}{2} V_\rho \right)
\quad {\rm and}\quad
\left( \begin{array}{c} \omega_{K^0}(\veck) \\ 
\omega_{\k0bar}(\veck) \end{array} \right) = 
\sqrt{m^\star_K + \vec{k}^2} \pm \left( V_\omega - \frac{1}{2} V_\rho \right). 
\label{dispresion}
\ene
This is equivalent to the dispersion relation which is given by the 
gauge invariant effective Lagrangian density at the hadronic 
level~\cite{schaff2,dai}: 
\bge
{\cal L} =  
\left[ \left(\partial_\mu + ig^K_\omega \omega_\mu 
+ ig^K_\rho \frac{\tau_3}{2} \rho_\mu \right) K \right]^\dagger
\left[ \left(\partial^\mu + ig^K_\omega \omega^\mu 
+ ig^K_\rho \frac{\tau_3}{2} \rho^\mu \right) K \right]
- m_K^{\star 2} \kbar K + {\cal L_{\rm matter}}\;,
\label{lagrangian}
\ene
where $K = \left( \begin{array}{c} K^+\\ K^0 \end{array} \right)$ is the 
second quantized kaon field with $\kbar = K^\dagger$, and 
${\cal L_{\rm matter}}$ is the Lagrangian density of a nuclear 
system~\cite{finite0,finite1,finite2}. 
In our approach, the effective mass, $m^\star_K$, is calculated 
using the MIT bag model, Eq.~(\ref{mass}) and the result is very 
well approximated by Eq.~(\ref{appmass}).

\begin{figure}[hbt]
\begin{center}
\epsfig{file=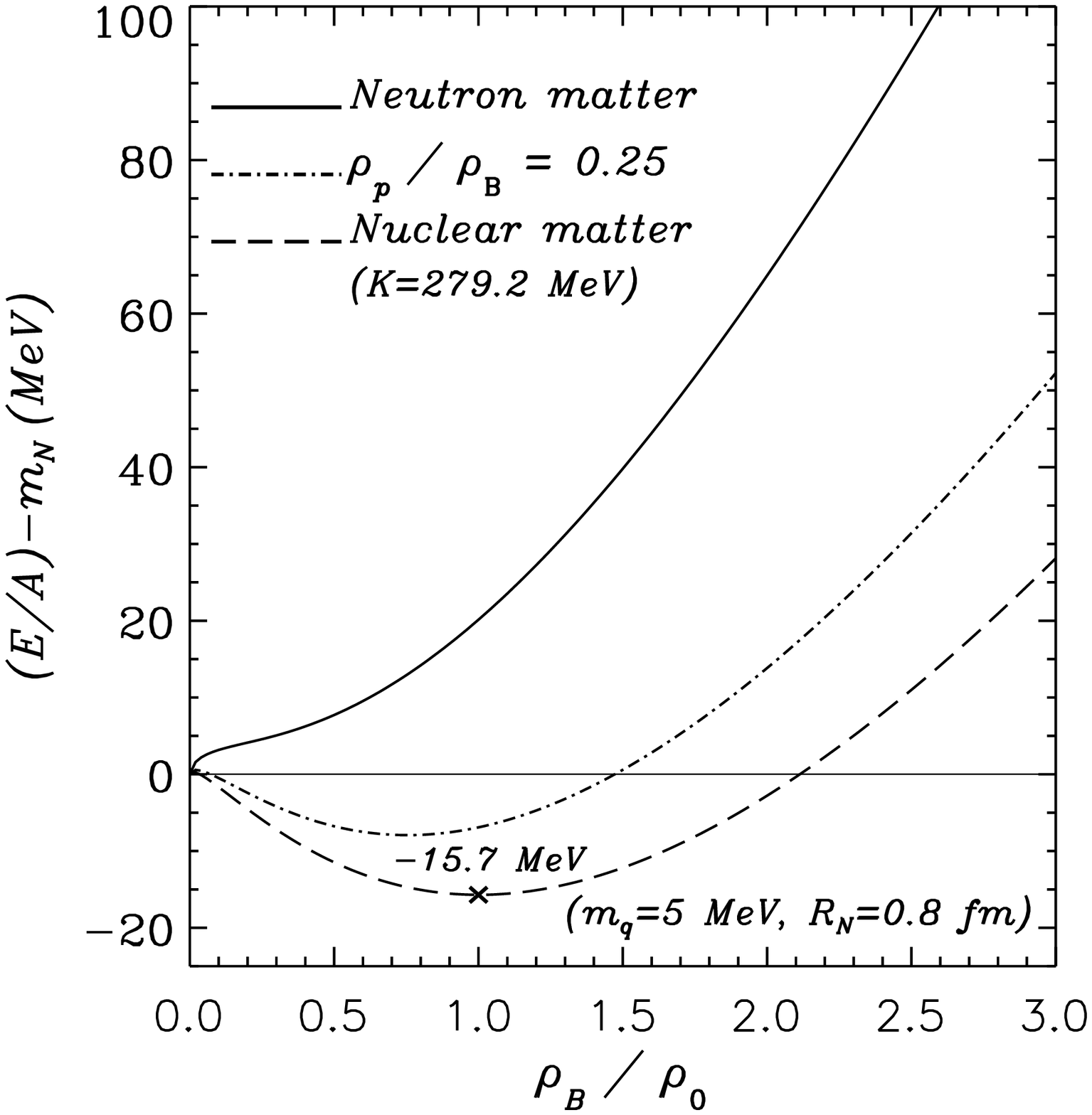,height=14cm}
\caption{Binding energy per nucleon for matter with 
different proton fractions. Neutron matter 
and nuclear matter denote matter with proton fractions 
0 and 0.5, respectively. $K = 279.2$ MeV is the 
value obtained for the nuclear incompressibility in the present model.} 
\label{neuteqs}
\end{center}
\end{figure}

In Fig.~\ref{neuteqs}, we show the binding energy per nucleon, 
$(E/A) - m_N$, versus baryon density.
Neutron matter and nuclear matter in Fig.~\ref{neuteqs} denote 
matter with proton fractions 0 and 0.5 
(symmetric matter), respectively. 
This notation will be used hereafter. The coupling constants, 
$g^N_\sigma, g^N_\omega$ and $g^N_\rho$ are determined so as to reproduce 
the saturation properties of symmetric nuclear matter 
at normal nuclear density, $\rho_0$, namely, the binding energy per 
nucleon, $- 15.7$ MeV, and bulk symmetry energy, 35.0 MeV. 
The values for the coupling constants determined in this way 
are given in Table~\ref{coupling}. Using these parameters, the nuclear 
incompressibility, $K$, is calculated to be $K = 279.2$ MeV, 
which is well within the empirically required range.
%

%
\begin{table}[htbp]
\begin{center}
\caption{
Values of the coupling constants determined required to reproduce the 
saturation properties of symmetric nuclear matter at normal nuclear 
density, $\rho_0 = 0.15$ fm$^{-3}$. For the relation between 
the coupling constants, $g^q_\sigma$ and $g^N_\sigma$, or, the origin 
of the constant factor between them, see
Refs.~\protect\cite{finite0,finite1}. 
}
\label{coupling}
\vspace{1em}
\begin{tabular}[hbtp]{|c|c|c|c|}
\hline
$g^q_\sigma = g^N_\sigma/(3 \times 0.483)$ 
&$(g^N_\sigma)^2/4 \pi$ &$(g^N_\omega)^2/4 \pi = 
(3 g^q_\omega)^2/4 \pi$ &($g^N_\rho)^2/ 4 \pi = (g^q_\rho)^2/4 \pi$\\
\hline
5.69 &5.39 &5.30 &6.93\\
\hline
\end{tabular}
\end{center}
\end{table}
%
%

\begin{figure}[hbt]
\begin{center}
\epsfig{file=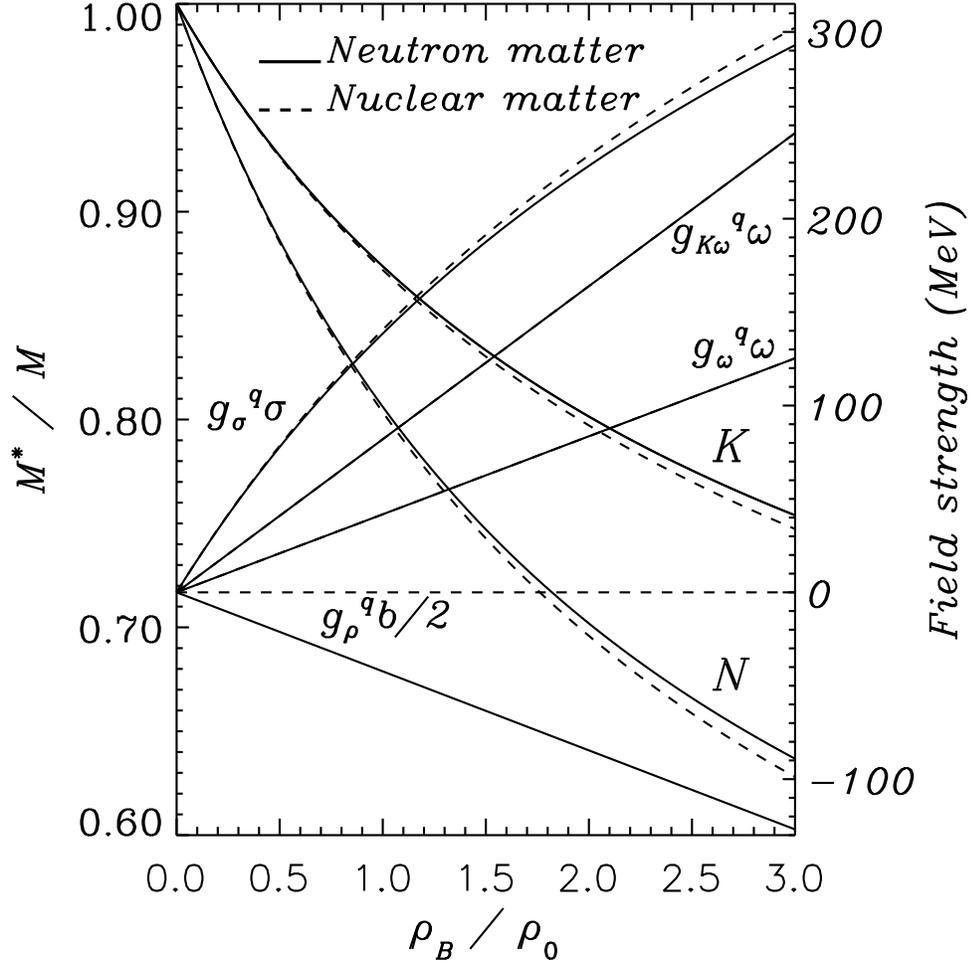,height=14cm}
\caption{Effective masses of the nucleon, $K$ (and $\kbar$) meson 
and the mean field potentials for the nonstrange quarks and antiquarks. 
$g_{K \omega}^q \omega$ is the $\omega$ meson mean field 
potential for the  
nonstrange quarks and antiquarks in the $K$ and $\kbar$ mesons. }
\label{neutmass}
\end{center}
\end{figure}

In Fig.~\ref{neutmass} we show the effective masses of the 
$K$ and $\kbar$ mesons and the nucleon, 
as well as the scalar and vector mean field potentials. 
{}For the present we have omitted any effects of hyperons
in the background nuclear medium. We observe that neither 
the effective mass of the $K$ and $\kbar$ mesons, 
$m^\star_K$, nor that of the nucleon 
decreases linearly as the density increases. 
This feature may be ascribed to the quark structure in the present approach, 
and clearly differs from the linear decrease in  
QHD~\cite{qhd}.  
This behaviour can be also understood from the 
parametrization, Eq.~(\ref{appmass}).  
$g_{K \omega}^q \omega$ in Fig.~\ref{neutmass} denotes the vector 
potential for the nonstrange quarks and antiquarks 
in the $K$ and $\kbar$ mesons, 
and we will explain this below, in connection with 
the $K^+ N$ potential.

It is known empirically that the $K^+ N$ potential is 
slightly repulsive if one wants to be consistent with the $K^+ N$ scattering 
length, and the corresponding value at $\rho_B = 0.16$ fm$^{-3}$ is 
estimated to be about 20 MeV~\cite{likpot}. On the other hand, 
the present model gives a very slightly attractive  
$K^+ N$ potential. 
We believe that this minor shortcoming has its origin in the deficiencies 
of the bag model in dealing with the Goldstone nature of the $K$ and 
$\kbar$ mesons.
As a phenomenological means of compensation for this 
we rescale the coupling constant, $g^q_\omega$, 
to reproduce the $K^+ N$ potential, $+ 20$ MeV, 
at $\rho_B = 0.16$ fm$^{-3}$.
That is, we use, $g^q_{K \omega} = 1.4^2 \times g^q_\omega$, for the  
$\omega$ meson coupling constant to the nonstrange quark 
(and antiquark) in the $K$ and $\kbar$ mesons. Note that the coupling 
constant, $g^q_{K \omega}$, is the only parameter adjusted in the 
present study. None of our qualitative conclusions would be altered if 
we did not make this adjustment.
The $\omega$ mean field vector potential calculated using 
this rescaled coupling constant
is denoted by $g^q_{K \omega} \omega$ in Fig.~\ref{neutmass}. 

\begin{figure}[hbt]
\begin{center}
\epsfig{file=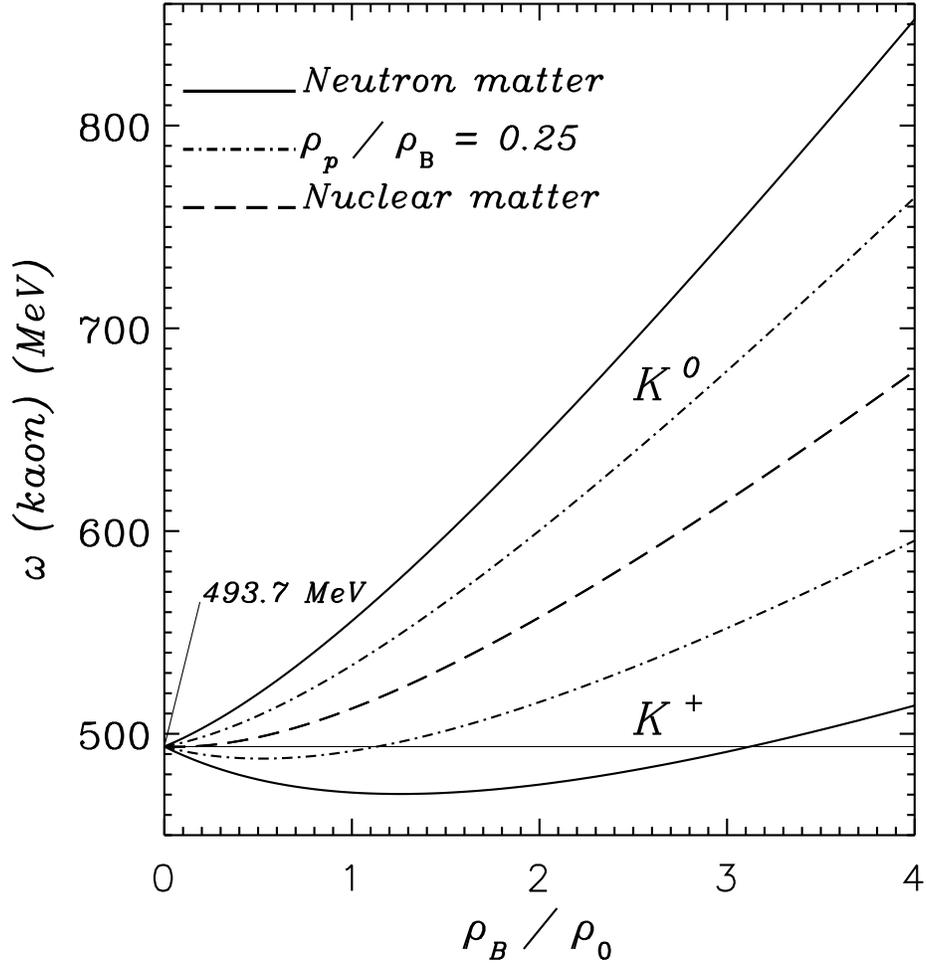,height=14cm}
\caption{Kaon excitation energies at zero momentum.} 
\label{neutkpot1}
\end{center}
\end{figure}

In Fig.~\ref{neutkpot1}, we show the calculated kaon excitation energies 
at zero momentum versus the baryon density.
It is interesting to notice that although the excitation energies for 
the isodoublet members, $K^+$ and $K^0$, are degenerate in 
symmetric nuclear matter, this is no longer true in asymmetric 
nuclear matter. This is a consequence of the $\rho$ meson 
which couples to the nonstrange quarks (antiquarks) in the kaon (antikaon).

\begin{figure}[hbt]
\begin{center}
\epsfig{file=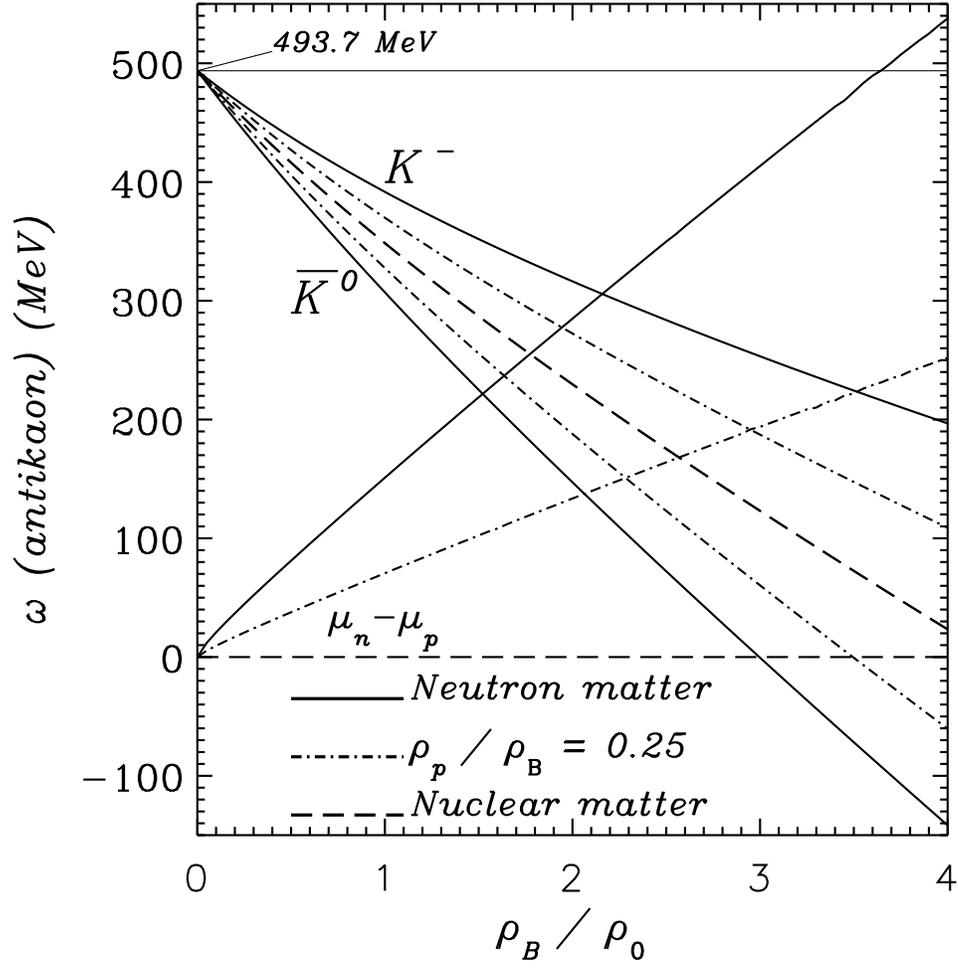,height=14cm}
\caption{Antikaon excitation energies at zero momentum, and 
the difference of the calculated chemical potentials for the neutron and 
proton, $\mu_n - \mu_p$.}
\label{neutkpot2}
\end{center}
\end{figure}

Next, we show the antikaon excitation energies in Fig.~\ref{neutkpot2},  
as well as the difference of the calculated 
chemical potentials for the neutron and proton, $\mu_n - \mu_p$, 
which is calculated by
\bge
\mu_n - \mu_p \simeq \left. \frac{E(Z,N)}{\partial N}\right|_Z
                 -   \left. \frac{E(Z,N)}{\partial Z}\right|_N 
= - \frac{\partial}{\partial x} \left( \frac{E(Z,N)}{A} \right) ,
\label{chemical}
\ene
where $Z$, $N$ ($A = Z + N$) and $E(Z,N)/A$ are, respectively, the 
proton number, neutron number and the total energy per nucleon with 
proton fraction, $x \equiv Z/A$.
Because we have not included any effects 
which are expected to lower the values of $\mu_n - \mu_p$, or  
raise the critical density for the onset of $K^-$ meson condensation 
(such as hyperons or muons~\cite{kno,schaff2,glen},
the $\delta$ meson~\cite{dai}, non-zero momentum for the  
$K^-$ mesons due to the thermal fluctuations or short-range correlations), 
the critical density found for each case  
may be regarded as a lower limit.
Again the excitation energies for 
the isodoublet members, $K^-$ and $\k0bar$, are no longer degenerate 
in asymmetric nuclear matter.
In particular, the excitation energy for the 
$K^-$ meson at a fixed density increases as the neutron fraction increases. 
Thus, the $\rho$ meson plays a role in making $K^-$ 
meson condensation 
less favorable to occur in a matter with a larger neutron excess. 
This effect of the $\rho$ field on the $K^-$ meson, certainly 
should be taken into account when one studies $K^-$-condensation 
and its effect on the properties of neutron stars.

If we use the parametrization of Eq.~(\ref{appmass}), together with 
the explicit expressions for the vector mean fields using 
the isoscalar ($\rho_B = \rho_p +\rho_n$) 
and isovector ($\rho_3 \equiv \rho_p - \rho_n$) 
baryon densities, the excitation energy for the $K^-$ meson 
at zero momentum, $\omega_{K^-}$, can be expressed as: 
\bg
\omega_{K^-} &\equiv& m^\star_K - V^K_\omega - \frac{1}{2} V_\rho, \nn \\
&\simeq& m_K - g^K_\sigma 
\left[ 1 - \frac{a_K}{2} (g^N_\sigma \sigma) \right] \sigma 
- \frac{g^K_\omega g^N_\omega}{m^2_\omega} \rho_B 
- \frac{g^K_\rho g^N_\rho}{4 m^2_\rho}  \rho_3,
\label{omega}
\en
where $g^K_\sigma = \frac{1}{3} g^N_\sigma$, 
$g^K_\omega = 1.4^2 \times g^q_\omega = 1.4^2 \times \frac{1}{3} g^N_\omega$ 
and $g^K_\rho = g^q_\rho = g^N_\rho$. 
For a rough estimate of the $K^-$ excitation energy 
up to $\rho_B \sim \rho_0$,  
one can use the approximate value for the scalar field 
$g^N_\sigma \sigma \simeq 200 \rho_B/\rho_0$ (MeV).

In summary, we have studied the properties of kaon and antikaon in 
nuclear matter, using the QMC model for the first time. 
Although the model should eventually incorporate chiral symmetry 
in order to treat the kaon and antikaon as pseudo-Goldstone bosons, 
our present emphasis was on the role of the $\rho$ meson 
in an asymmetric nuclear medium. In particular,  
in matter with a neutron excess, or with a negative isovector density, 
the $\rho$ meson induces a repulsive potential for the 
$K^-$ meson. 
This effect should certainly be taken into 
account in investigations of the properties of neutron stars, and kaon flow
in heavy ion collisions, where it has so far been omitted~\cite{liflow}.
Indeed, it may be possible to test our estimate of this effect 
by calculating the $\k0bar$ and
$K^-$ flow in heavy ion collisions.
In the present study, we have not included any effects 
which are expected to lower the chemical potential of the electron 
such as a non-zero hyperon density~\cite{kno,schaff2,glen}.
For a more realistic study, it will be necessary to include self-consistently 
the effect of the hyperons in calculating  
scalar and vector fields. In that case, it is possible that the quark  
structure of the hadrons, which appears mainly in a non-linear  
variation of their effective masses, may give nontrivial effects. 
In particular, our further interest is 
whether the QMC model can avoid the negative effective mass problem 
for the nucleon, that was discussed by Schaffner 
and Mishustin~\cite{schaff2}.
\vspace{1cm}

\noindent
The authors would like to thank G.Q. Li, T. Tatsumi and A.G. Williams for 
helpful discussions. This work is supported by the Australian 
Research Council and the Japan Society for the Promotion of Science.

%

%
%
%
\newpage
%
%
%
\newpage
%
%
%
%
%
%
\newpage
%
%
%
\newpage
%
%
\end{document}